\documentclass[nohyperref]{article}

\usepackage{microtype}
\usepackage{graphicx}
\usepackage{subfigure}
\usepackage{booktabs} 
\usepackage{hyperref}
\usepackage[accepted]{icml2022}
\usepackage{amsmath}
\usepackage{amssymb}
\usepackage{mathtools}
\usepackage{amsthm}


\theoremstyle{plain}

\theoremstyle{definition}

\theoremstyle{remark}

\definecolor{es-blue}{rgb}{0,0.4,0.8}

\icmltitlerunning{Strong Lensing Source Reconstruction Using Continuous Neural Fields}

\begin{document}

\twocolumn[
\hfill MIT-CTP/5444
\icmltitle{Strong Lensing Source Reconstruction Using Continuous Neural Fields}

\begin{icmlauthorlist}

\icmlauthor{Siddharth Mishra-Sharma}{iaifi,ctp,harvardphys}
\icmlauthor{Ge Yang}{iaifi,csail}
\end{icmlauthorlist}

\icmlaffiliation{iaifi}{NSF Institute of AI and Fundamental Interactions (IAIFI)}
\icmlaffiliation{csail}{Computer Science and Aritificial Intelligence (CSAIL), MIT}
\icmlaffiliation{ctp}{Center for Theoretical Physics, Massachusetts Institute of Technology, Cambridge, MA 02139, USA}
\icmlaffiliation{harvardphys}{Department of Physics, Harvard University, Cambridge, MA 02138, USA}
\icmlcorrespondingauthor{Siddharth Mishra-Sharma}{smsharma@mit.edu}
\icmlcorrespondingauthor{Ge Yang}{geyang@csail.mit.edu}

\icmlkeywords{volumetric rendering, analysis by synthesis, gravitational lensing, implicit neural representations, continuous neural fields, computer vision, astrophysics, dark matter}

\vskip 0.3in
]

 \printAffiliationsAndNotice{} 

\begin{abstract}
From the nature of dark matter to the rate of expansion of our Universe, observations of distant galaxies distorted through strong gravitational lensing have the potential to answer some of the major open questions in astrophysics. Modeling galaxy-galaxy strong lensing observations presents a number of challenges as the exact configuration of both the background source and foreground lens galaxy is unknown. A timely call, prompted by a number of upcoming surveys anticipating high-resolution lensing images, demands methods that can efficiently model lenses at their full complexity. In this work, we introduce a method that uses continuous neural fields to non-parametrically reconstruct the complex morphology of a source galaxy while simultaneously inferring a distribution over foreground lens galaxy configurations. We demonstrate the efficacy of our method through experiments on simulated data targeting high-resolution lensing images similar to those anticipated in near-future astrophysical surveys. 
\end{abstract}

\section{Introduction}
\label{sec:intro}

According to general relativity, the warping of space-time in the vicinity of a massive celestial body can distort the path of light rays traversing close-by. This phenomenon, called \textit{gravitational lensing}~\cite{Einstein1936LENS}, is a powerful probe of the distribution of structures in our Universe. In the \emph{strong lensing} regime, multiple highly magnified images of background luminous sources can be observed.  This is a versatile astrophysical laboratory that has been used to characterize distant galaxies~\cite{wuyts2012constraints,yuan2017most}, constrain the abundance of dark matter substructures~\cite{dalal2002direct,vegetti2014inference,hezaveh2016detection,gilman2019warm,hsueh2020sharp,csengul2021substructure,meneghetti2020excess}, and provide percent-level estimates of the expansion rate of the Universe~\cite{birrer2020tdcosmo}.

A particular challenge with fully exploiting observations of galaxy-galaxy strong gravitational lenses --- where an extended background source is lensed by a foreground galaxy~---~is that of accounting for the complex morphologies of lensed galaxies. Although sources in low-resolution images can be adequately modeled using phenomenological parameterizations such as one or several S\'{e}rsic profiles~\citep{sersic1963influence}, this approach is inadequate for modeling higher-fidelity lensing observations such as those from ongoing, upcoming, and proposed telescopes like the \emph{Hubble} Space Telescope (HST), JWST, \emph{Euclid}, and the Extremely Large Telescope (ELT). The development of new methods is especially timely, given the large number of high-resolution lenses that are expected to be imaged by next-generation cosmological surveys~\citep{collett2015population} and their potential to weigh in on the nature of dark matter~\cite{simon2019testing}.

A number of methods have been proposed that go beyond using simple parameterization for the background galaxies. They include regularized semi-linear inversion~\citep{warren2003semilinear}, the use of truncated bases like shapelets~\citep{birrer2015gravitational,birrer2018lenstronomy} or wavelets~\citep{galan2021slitronomy}, and the use of adaptive source-plane pixelizations~\citep{vegetti2009bayesian,nightingale2018autolens}. More recently, deep learning techniques, for example variational autoencoders~\citep{chianese2020differentiable}, recurrent inference machines~\citep{morningstar2019data}, and Gaussian process-inspired inference~\citep{karchev2022strong} have also shown promise. 
Downstream analyses often benefit from, and sometimes require, a probabilistic treatment of both the source and lens. Using variational methods, \citealt{karchev2022strong} demonstrated the ability to recover the posterior distribution over both the lens parameter and a complex source configuration.

\begin{figure*}[!t]
\centering
\includegraphics[width=0.99\textwidth]{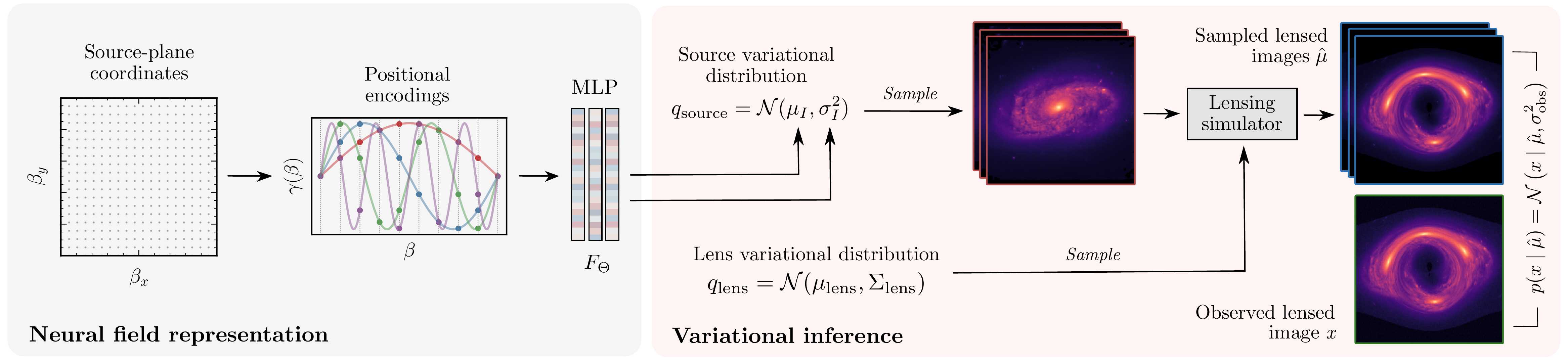} \\
\caption{A schematic overview of the method used in this work.}\label{fig:figure}
\end{figure*}

In this paper, we explore a complementary approach for probabilistic reconstruction of complex morphologies of strongly lensed galaxies from high-resolution images while simultaneously recovering the lens mass distribution. Our key contribution is to treat the source-light distribution as a continuous neural field, optimized using gradient-based variational inference. Together with recovering a posterior distribution over the lens model parameters, our approach is able to infer high-resolution images of the source galaxy in its full complexity. In the context of strong lensing, neural fields have previously been proposed for modeling the density profile of the smooth lens galaxy~\cite{biggio2021ml4ps}.

\section{Methodology}
\label{sec:method}

Our proposal takes the ``analysis by synthesis'' approach to conduct probabilistic inference on both the source image and the lens configuration using a differentiable lensing model. We describe the three key components of this generative pipeline below: \emph{(1)}~the differentiable lensing model used for rendering and the relevant strong-lensing physics, \emph{(2)}~the continuous neural representation of the source, and \emph{(3)}~the variational inference procedure used to simultaneously infer the lens parameter and source posteriors.

\subsection{Strong Lensing and the Synthesis Model}

At the heart of our approach is a differentiable renderer that takes the source and lens configurations, and outputs modeled lensing observations. The simulator we use is a modified version of \texttt{gigalens}~\cite{gu2022giga}, written in \texttt{Jax}~\cite{jax2018github}. 

The position of the source in the lens plane  $\theta$ can be evaluated using the lens equation, 
$\beta = \theta - \phi(\theta)$.
Here $\beta$ is the position in the source plane and $\phi(\theta)$ is the deflection vector, given by the gradient of the projected gravitational potential $\psi_\mathrm{G}(\theta)$ of the lens,
$\phi(\theta) = \nabla\psi_\mathrm{G}(\theta)$. The lens equation relates the lens-plane coordinates back to those in the source-plane. Given an extended source light profile $f_\mathrm{s}$ and a lensing mass distribution, we can reconstruct the lens-plane observation $f_\mathrm{s}'$ by evaluating the source light on the lens plane, $f_{\mathrm{s}}^{\prime}({\theta})=f_{\mathrm{s}}({\theta}-{\phi}({\theta}))$. We refer to, e.g., \citet{treu2010strong} for additional details of the strong lensing formalism.

The main lens deflector is modeled using the commonly employed \textit{Singular Isothermal Ellipsoid} (SIE) parameterization~\cite{1994A&A...284..285K,treu2010strong}. The deflection vector field in this case is given in terms of angular coordinates $\theta_x$ and $\theta_y$ via (see e.g., ~\citealt{keeton2001catalog}) 
\begin{equation}
    \begin{aligned}
        &\phi_{\mathrm{lens}, x}=\frac{\theta_{\mathrm{E}} q}{\sqrt{1-q^{2}}} \tan ^{-1}\left[\frac{\sqrt{1-q^{2}} \theta_{x}}{\chi}\right] \\
        &\phi_{\mathrm{lens}, y}=\frac{\theta_{\mathrm{E}} q}{\sqrt{1-q^{2}}} \tanh ^{-1}\left[\frac{\sqrt{1-q^{2}} \theta_{y}}{\chi+q^{2}}\right],
    \end{aligned}
\end{equation}
where $q$ is the axis ratio ($q=1$ corresponding to a spherical lens), $\chi \equiv \sqrt{\theta_{x}^{2} q^{2}+\theta_{y}^{2}}$, and $\theta_\mathrm{E}$ is the Einstein radius denoting the characteristic lensing scale.

The lens orientation is specified in terms of eccentricities $\left(\epsilon_{1}, \epsilon_{2}\right)=\frac{1-q}{1+q}\left(\cos (2 \psi), \sin (2 \psi)\right)$, where $\psi$ is a rotation angle. The large-scale lensing effect of the local environment is included through an external shear component $\{\gamma_1, \gamma_2\}$ with deflection vector $\left(\phi_{\mathrm{ext},x},\phi_{\mathrm{ext},y}\right) = \left(\gamma_1\theta_x + \gamma_2\theta_y, \gamma_2\theta_x - \gamma_1\theta_y\right)$. 

After allowing for an overall offset $\left(\theta_{x,0}, \theta_{y,0}\right)$ between the lens and source center lines of sight, our lens mass model consists of 7 parameters $\{\theta_\mathrm{E}, \theta_{x,0}, \theta_{y,0}, \epsilon_1, \epsilon_2, \gamma_1, \gamma_2\}$. 
We include the full list of lens model parameters and their assumed prior distributions in Tab.~\ref{tab:lens_params}.
\begin{table}[t]
\centering
\caption{List of parameters used in the lens model and their corresponding assumed prior distributions.}
\begin{tabular}[t]{lcc}
\toprule
Parameter & Symbol & Prior\\
\midrule
Einstein radius & \(\theta_{\mathrm E}\) & $\mathcal U(1'', 2'')$ \\
Source-lens offset &\(\theta_{x,0}, \theta_{y,0}\) & $\mathcal U(-0.5'', 0.5'')$ \\
Eccentricities & \(\epsilon_1, \epsilon_2\)& $\mathcal N(0, 0.3)$\\
External shear &\(\gamma_1, \gamma_2\) & $\mathcal N(0, 0.05)$ \\
\bottomrule
\end{tabular}
\label{tab:lens_params}
\end{table}

\subsection{Continuous Neural Representation of the Source}\label{sec:continuous}

\begin{figure*}[!t]
\centering
\includegraphics[width=0.99\textwidth]{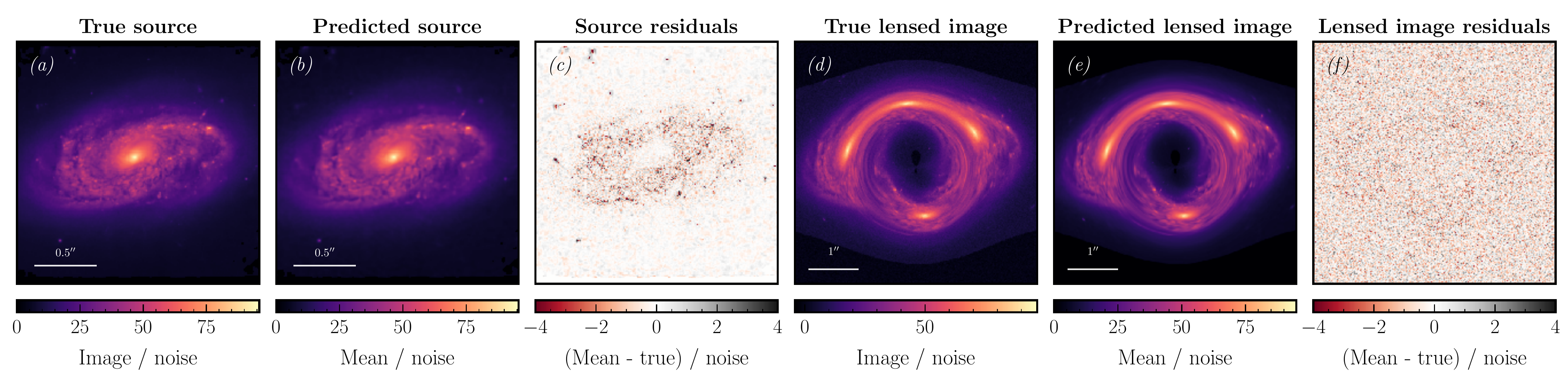} \\
\caption{Results of reconstructing the source through our coordinate-based neural network pipeline using a mock lensed image of galaxy NGC2906. The \emph{(a)} true source, \emph{(b)}  reconstructed source mean, \emph{(c)}  reconstructed mean minus true source residuals,  \emph{(d)} true lensed image,  \emph{(e)} reconstructed lensed image, and  \emph{(f)} reconstructed mean minus true lensed image residuals are shown. All images are normalized by the observation noise.}\label{fig:reconstruction}
\end{figure*}
In our framework, the source-light distribution is modeled using a continuous neural field  $F_\Theta: \mathbb R^2 \mapsto \mathbb R^2$, which takes the source-plane coordinates $\{\beta_x, \beta_y\}$ in \(\mathbb R^2\) as input and outputs the mean and variance of the light intensity, modeled as a Gaussian distribution $I({x, y}) \sim \mathcal N\left(\mu_I({x, y}), \sigma^2_I({x, y})\right)$.

Based on the universal function approximation theorem~\cite{hornik1989multilayer}, a na\"ive choice is to parameterize \(F_\Theta\) as a vanilla neural network. Although the theorem guarantees the existence of good approximations, it makes no claims of how easy or difficult it would be to attain them. Empirical analyses uncovered an intrinsic learning bias for standard network architectures to favor fitting simpler, low-spatial frequency components of the target function under gradient descent~\cite{zhang2021understanding,arpit2017closer,ulyanov2018deep,rahaman2019spectral}.  Using new tools such as the \textit{neural tangent kernel} (NTK,~\citealt{jacot2018neural}), recent progress in deep learning theory describe this phenomena as a ``spectral bias'' that slows down convergence over high-frequency components of the target function exponentially, which often result in poorly-fit models~\cite{rahaman2019spectral,cao2019towards}.

This problem can be addressed by adding positional encodings, now ubiquitously used in natural language processing~\cite{gehring2017convolutional,vaswani2017attention,devlin2018bert}. 
For an input \(\beta\in \mathbb R\), positional encodings lift the input coordinates \(\beta\) into a higher-dimensional spectral domain with \(2 \times L_\mathrm{max}\) dimensions per input dimension:
\begin{align}
\begin{split}
\gamma(\beta)=\big[&\sin (\beta), \cos (\beta), \sin(2 \beta), \cos(2 \beta), \ldots  \\
&\sin \left(2^{L_\mathrm{max}-1} \beta\right), \cos \left(2^{L_\mathrm{max}-1} \beta\right)\big]^{\mathrm{T}},
\label{eq:pos_enc}
\end{split}
\end{align}
where $L_\mathrm{max}$ is the maximum bandwidth of the encoding. When applied to low-dimensional spaces such as image coordinates, such a scheme synthesizes high-fidelity reconstructions~\cite{mildenhall2020nerf}, and was most recently applied in the astrophysics domain for tomographic reconstruction of lensed black hole emission in interferometric observations~\citet{levis2022gravitationally}. 
We implement $F_\Theta$ as a four-layer multi-layer perceptron (MLP) with ReLU activations and hidden dimension \(256\). The intrinsic modeling bias of such networks is described in detail in~\cite{tancik2020fourier}.

Higher values of the bandwidth hyperparameter $L_\mathrm{max}$ encourage reconstructing higher-frequency features, but may lead to overfitting on noise. We found $L_\mathrm{max}=5$ to work well in our set-up. Alternative schemes for lifting the input coordinates such as random Fourier features may be advantageous~\cite{Rahimi2008kitchen}, and we discuss these in Sec.~\ref{sec:discussion} and in App.~\ref{app:pos_enc}.

\subsection{Variational Inference}

We use variational inference~\cite{jordan1999introduction,blei2017variational} to simultaneously fit for the joint distribution of the $n_\mathrm{lens} = 7$ lens parameters alongside the source intensity distribution on a specified grid on the source plane. The variational ansatz on the lens parameters is taken to be a multivariate Normal distribution $q_\mathrm{lens} = \mathcal N(\mu_\mathrm{lens}, \Sigma_\mathrm{lens})$, where $\mu_\mathrm{lens}$ is the mean vector and $\Sigma_\mathrm{lens}$ the parameter covariance matrix. We fit for the $n_\mathrm{lens} (n_\mathrm{lens} + 1) / 2$-parameter lower-triangular Cholesky factor $L$ of the covariance matrix, $\Sigma_\mathrm{lens} \equiv LL^T$, ensuring positivity of the diagonal component through the projection $\operatorname{Softplus}(\cdot) \equiv \log\left(1 + e^{(\cdot)}\right)$.

The source on the other hand is modeled using a diagonal Normal distribution $q_\mathrm{source} = \mathcal N(\mu_I, \sigma_I^2)$, with the means and variances evaluated on the source grid as the outputs of the neural network $F_\Theta$. In this proof-of-principle exposition we do not model correlations between the lens and source parameters, leaving their inclusion to future work. In the following, the lens and source parameters (SIE parameters and intensities on a grid, respectively) are jointly denoted by $\vartheta$.

Given a pixelized lensed image $x$, the source and lens variational distributions, parameterized by $\phi$ and jointly denoted $q_\phi(\vartheta)$, are obtained by minimizing the reverse KL-divergence between the true and variational posterior distributions, $D_\mathrm{KL}\left(q_\phi\mid\mid p(\vartheta\mid x)\right)$. This is done by maximizing the model log-evidence $\log p(x)$ or, in practice using the tractable evidence lower bound (ELBO), 
\begin{equation}
\operatorname{ELBO} \equiv \mathbb E_{\vartheta\sim q_\phi(\vartheta)}\left[\log p(x, \vartheta) - \log q_\phi(\vartheta)\right],
\end{equation}
as the optimization objective since $\log p(x) - \operatorname{ELBO} = D_\mathrm{KL}\left(q_\phi\mid\mid p(\vartheta\mid x)\right)$. The expectation in the ELBO is taken through Monte Carlo sampling from the source and lens variational distributions at each optimization step. ELBO gradients are straightforwardly computed using the reparameterization trick~\cite{kingma2013auto} since all the distributions employed are reparameterizable. We emphasize that the model only has access to the \emph{lensed} image $x$ and not the ground-truth source.

Monte Carlo samples from the lens and source distribution are passed through the differentiable lensing pipeling to produce lensed image samples $\{\hat\mu(\vartheta)\}$. The joint likelihood $p(x,\vartheta) = p(x\mid\vartheta)\,p(\vartheta)$ is evaluated by combining the prior with the data likelihood, modeled as Normal; $p(x\mid\vartheta) = \mathcal N\left(x\mid \hat\mu(\vartheta), \sigma_\mathrm{obs}^2\right)$. Priors on the lens parameters are shown in Tab.~\ref{tab:lens_params}.

The variational parameters---the weights and biases of  $F_\Theta$ in the case of the source, and the mean and lower-triangular Cholesky matrix elements for the lens---are optimized using the AdamW optimizer~\cite{kingma2014adam,loshchilov2017decoupled} over 15,000 steps. The learning rate is increased from zero to $10^{-3}$ over 2000 steps, then decreased via cosine annealing over the course of optimization. Fitting a lensed image typically takes $\sim 20$ minutes on a single Nvidia V100 GPU. The pipeline is constructed using the neural network and optimization libraries \texttt{Flax}~\citep{flax2020github} and \texttt{Optax}~\citep{optax2020github}.

A schematic illustration of the high-level features of the method are shown in Fig.~\ref{fig:figure}.

\section{Experiments}
\label{sec:experiments}

We test our source and lens reconstruction pipeline using an image of galaxy NGC2906 imaged by the \emph{Hubble} Space Telescope\footnote{\url{https://esahubble.org/images/potw2015a/}} as the mock source. A lensed image of size $500\times500$ pixels is generated with image side 5 arcmin, corresponding to pixel size of 10 mas. A Gaussian point-spread function kernel is included with FWHM of 5 mas~\cite{simon2019testing}, although it has minimal effect given the pixel scale.  Uniform Gaussian noise is added to the pixels such that the mean signal-to-noise ratio around the lensed ring (defined as pixels that are at least 50\% as bright as the brightest pixel) is $\mathrm{SNR}\sim 60$. The lens parameters used to simulate the mock image are chosen as $\{\theta_\mathrm{E}, \theta_{x,0}, \theta_{y,0}, \epsilon_1, \epsilon_2, \gamma_1, \gamma_2\} = \{1.3'', 0.1'', 0.0'', 0.2, 0.1, -0.02, 0.02\}$.

\begin{figure*}[!ht]
\centering
\includegraphics[width=0.75\textwidth]{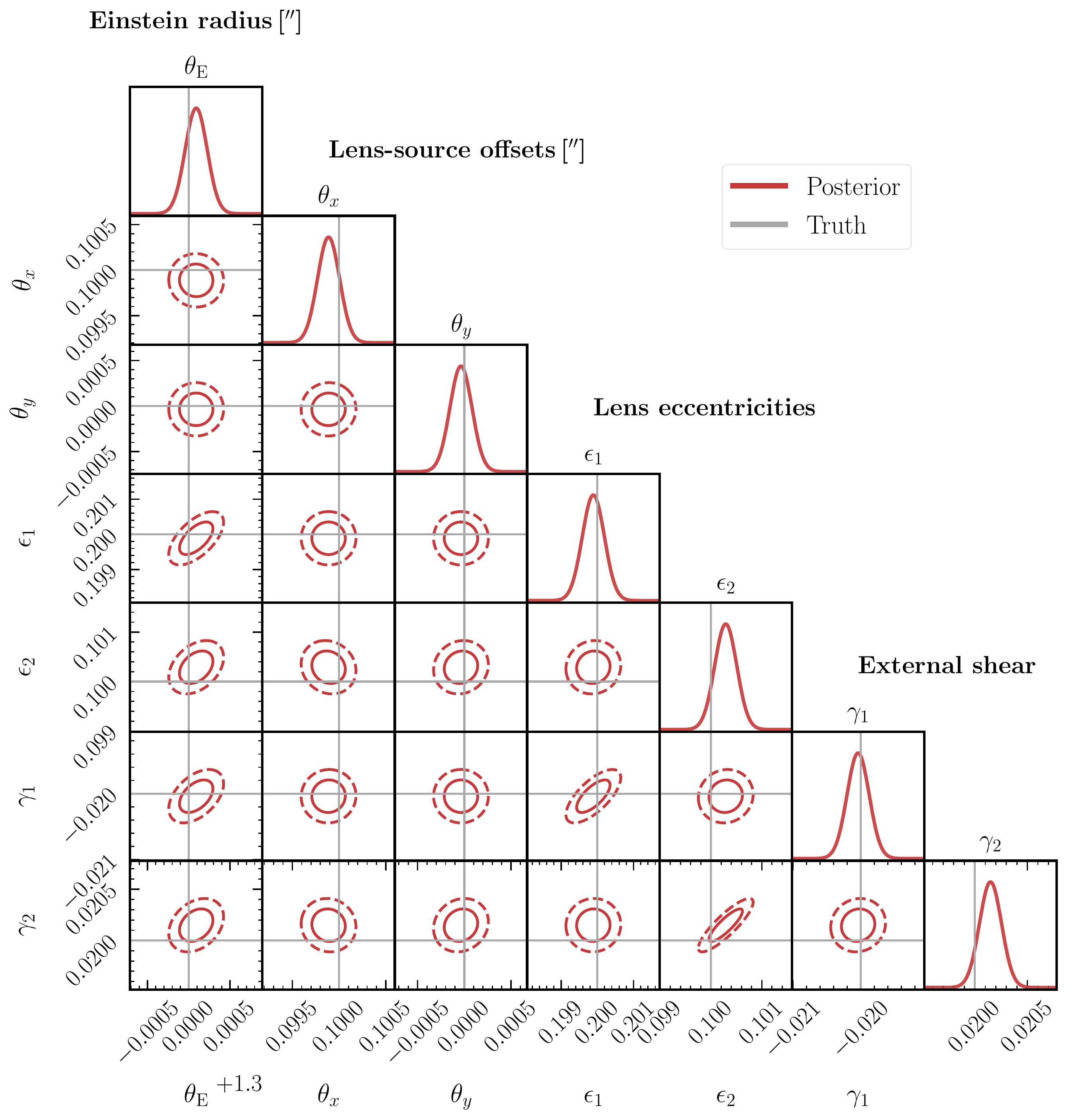}
\vspace{-1em}
\caption{Individual and joint marginal posteriors on the SIE lens galaxy and external shear parameters obtained in our experiment. The ellipses show the 1- and 2-$\sigma$ containment of the joint posteriors. True parameter points from which the lensed image was simulated are indicated with grey lines. Faithful recovery of the true parameters is seen.}\label{fig:joint_lens}
\vspace{-0.5em}
\end{figure*}
The results of our source and lensed image reconstruction are shown in Fig.~\ref{fig:reconstruction}, with all results normalized to the standard deviation of observation noise. The reconstructed source \emph{(b)} is seen to provide a visually good representation of the unlensed galaxy \emph{(a)}. The reconstructed lensed image \emph{(e)} again faithfully reproduces the observed lensed image \emph{(d)}, additionally corroborated by the normalized residuals \emph{(f)}. Some significant residual structures in the source plane are observed in \emph{(c)}; this can at least partially be attributed to a simplistic choice of source pixelization, discussed below. Example reconstructions for additional mock source and lens configurations are presented in App.~\ref{app:additional}.

Figure~\ref{fig:joint_lens} shows the 1-D and 2-D joint marginal posteriors inferred for the lens and external shear parameters in our experiment. 1-D marginals are shown as Gaussian profiles, while the 2-D joint distributions are represented through their 1- and 2-$\sigma$ containment regions. The true simulated parameters are indicated with grey lines. Successful recovery of the lens parameters within modeled uncertainty is seen in all cases.

\section{Discussion and Future Work}
\label{sec:discussion}

In this ongoing work, we have explored the use of continuous neural fields for modeling morphologically complex gravitationally lensed galaxies. Our method is able to provide a probabilistic reconstruction of lensed galaxies with complex small-scale structure within minutes of wall-clock time, while simultaneously inferring a posterior over the parameterized macroscopic lens deflector field. Although our results are encouraging, practical deployment of the method will require several extensions and modifications to the framework. We conclude by discussing these extensions alongside caveats associated with the proposed method.

In our method, the prior over the source reconstruction is implicit, and it occurs through two separate mechanisms. First, it is affected by the bandwidth limit used in the positional encoding. Second, a low-frequency bias occurs through early stopping of the optimization process, which prevents the neural network from fitting higher-frequency features and overfitting on noise. Although the dynamics associated with with this phenomenon have been well-studied~\citep{roberts2022deep-learning}, a dedicated understanding in the context of our specific domain could inform more principled regularization for a given lens observation configuration. Empirically, we found that excessive optimization leads to overfitting on noisy images, which in turn can produce systematic biases in the reconstruction of lens galaxy parameters.

The positional encoding used in this paper is non-isotropic under rotation, because it processes the two axes independently from each other. An isotropic alternative is random Fourier features (RFF, see~\citet{Rahimi2008kitchen}), which samples a projection matrix from an isotropic distribution such as a multivariate Gaussian~\cite{yang2022overcoming}. Random Fourier features treat the input coordinates as full vectors during the spectral projection. The downside is due to its statistical nature, RFF require a larger number of features. The cut-off band-limit of both encodings can be controlled explicitly through the spectral mixture used to lift the input. In a learned variant of RFF, called the learned Fourier features, these weights are updated through gradient descent. This means more careful regularization of the cut-off bandwidth of the projection matrix is needed.

Modifications to the lensing pipeline can further improve the quality and robustness of reconstruction. For example, the current implementation glosses over subtleties associated with the source-plane pixelization. Instead, it models the pixelated data in the lens plane through individual points in the source plane. In addition, representing the source plane with the same resolution as the lens plane has been shown to be sub-optimal given the coordinate distortion inherent to the lensing process~\citep{warren2003semilinear}; the contribution of source-plane regions corresponding to larger magnification should be evaluated at a relatively finer resolution. Our {continuous}, rather than discretely-sampled, representation of the source should make our framework amenable to a more principled treatment of the the source-plane pixelization.

Finally, it is possibly to use more expressive variational distributions beyond the Gaussian ansatz and explicitly model covariances between the source galaxy and lens parameters. This could be accomplished, for example, by modeling the lens parameters using a normalizing flow conditioned on summary features extracted from the reconstructed source over the course of optimization. Having demonstrated the potential of the method in a proof-of-principle setting, we leave these extensions to future work.  

\section*{Software and Data}
Code for reproducing the results presented in this paper is available at \url{https://github.com/smsharma/lensing-neural-fields}. 
This research made use of the 
\texttt{Astropy}~\citep{astropy:2013, astropy:2018, astropy:2022},
\texttt{Flax}~\citep{flax2020github}, 
\texttt{gigalens}~\citep{gu2022giga}, 
\texttt{IPython}~\cite{PER-GRA:2007},
\texttt{Jax}~\cite{jax2018github,deepmind2020jax},
\texttt{jaxinterp2d}~\cite{jaxinterp2dgithub},
\texttt{Jupyter}~\cite{Kluyver2016JupyterN},
\texttt{lenstronomy}~\cite{birrer2018lenstronomy,Birrer2021},
\texttt{Matplotlib}~\cite{Hunter:2007},
\texttt{NumPy}~\cite{harris2020array},
\texttt{numpyro}~\cite{phan2019composable},
\texttt{Optax}~\citep{optax2020github},
\texttt{scikit-image}~\cite{scikit-image}, and
\texttt{SciPy}~\cite{2020SciPy-NMeth}
software packages.

\section*{Acknowledgements}

This work was performed in part at the Aspen Center for Physics, which is supported by National Science Foundation grant PHY-1607611.
This work is supported by the National Science Foundation under Cooperative Agreement PHY-2019786 (The NSF AI Institute for Artificial Intelligence and Fundamental Interactions, \url{http://iaifi.org/}).
This material is based upon work supported by the U.S. Department of Energy, Office of Science, Office of High Energy Physics of U.S. Department of Energy under grant Contract Number DE-SC0012567.
The computations in this paper were run on the FASRC Cannon cluster supported by the FAS Division of Science Research Computing Group at Harvard University.

\bibliography{lensing-neural-fields}
\bibliographystyle{icml2022}

\newpage
\appendix
\onecolumn

\section{Role of positional encodings}
\label{app:pos_enc}

\begin{figure*}[!ht]
\centering
\includegraphics[width=0.75\textwidth]{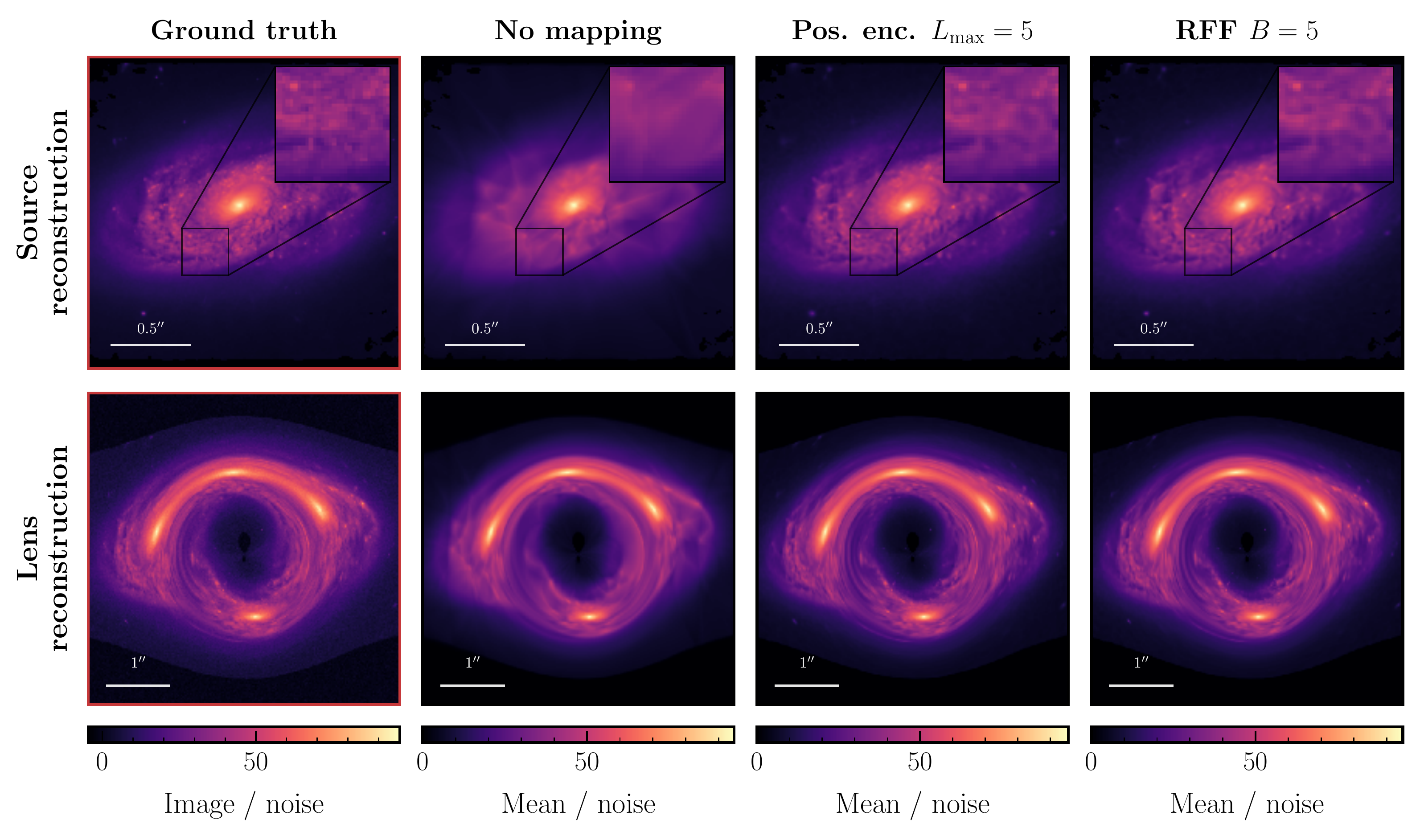}
\vspace{-1em}
\caption{Source (top row) and lens (bottom row) reconstruction using variations on mapping the input coordinates to a higher-dimensional feature space. The ground truth (first column), no coordinate mapping (second column), positional encodings (third column; baseline), and Gaussian random Fourier features (last column) are shown. Poor reconstruction of small-scale source galaxy features is seen when not employing an input mapping.}
\label{fig:encodings}
\vspace{-0.5em}
\end{figure*}

Mapping the input source coordinates into a specific higher-dimensional feature space encourages effective reconstruction of small-scale (high-frequency) features in the source galaxy. This is demonstrated in Fig.~\ref{fig:encodings}, where we show the baseline reconstructed source along with a zoom-in inset (top row) and lens (bottom row) for various mapping configurations. The ground truths (first column), results with no coordinate mappings (second column), and standard sinusoidal positional encodings (Eq.~\eqref{eq:pos_enc} with bandwidth hyperparameter set to $L_\mathrm{max} = 5$; third column) are shown. It can be seen that small-scale galaxy features are better reconstructed when using positional encodings, whereas these appear washed out without the additional encoding step.

Additionally, we show results using a Gaussian random Fourier feature mapping~\cite{tancik2020fourier,rahimi2007random,yang2022overcoming}. Here, the source coordinates $\beta$ are lifted through the mapping $\mathrm{RFF}(\beta)_i =\sin (w_{ij}\beta_j + b_i)$ where the weights and biases are drawn as $w_{ij}\sim\mathcal N\left(0, (B/2)^2\right)$ and $b_i \sim \mathcal U(-1, 1)$ at initialization. The last column of Fig.~\ref{fig:encodings} shows results when using this mapping, setting the bandwidth hyperparameter $B=5$. Similar performance in reconstructing small-scale galaxy features to that obtained using the standard positional encoding scheme can be seen in this case.

\section{Additional examples}
\label{app:additional}

\begin{figure*}[t]
\centering
\includegraphics[width=0.99\textwidth]{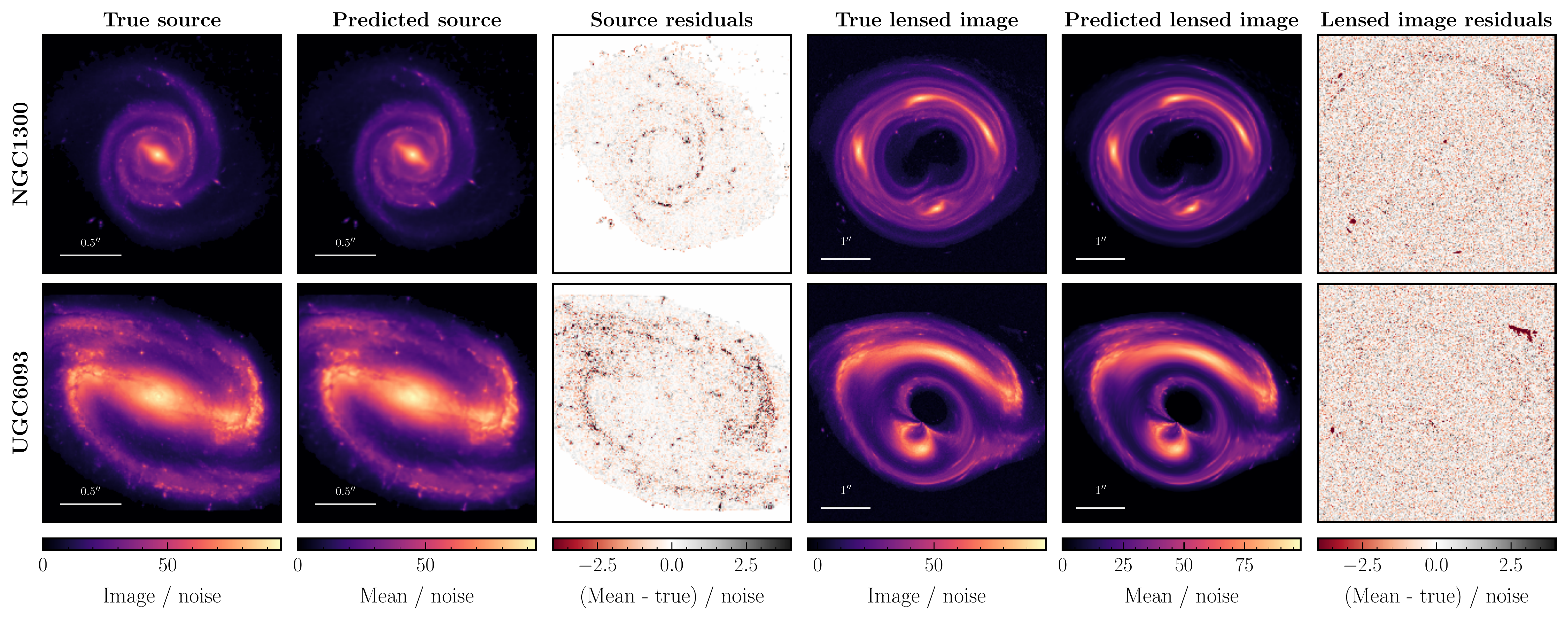}
\caption{Reconstruction results using additional source and lens configurations, in the same format as Fig.~\ref{fig:reconstruction}.}
\label{fig:reconstruction_additional}
\end{figure*}

In Fig.~\ref{fig:reconstruction_additional} we show additional source reconstruction results on mock images simulated using images of galaxies NGC1300\footnote{\url{https://esahubble.org/images/opo0501a/}} and UGC6309\footnote{\url{https://esahubble.org/images/potw1801a/}} taken by \emph{Hubble}. The same lens configuration as in the baseline example is used in the first case, while in the second case we set $\{\theta_\mathrm{E}, \theta_{x,0}, \theta_{y,0}, \epsilon_1, \epsilon_2, \gamma_1, \gamma_2\} = \{1.1'', 0.4'', -0.1'', 0.1, 0.15, -0.03, 0.04\}$.

\end{document}